\documentclass[aip,twocolumn,amsmath,reprint,amssymb,graphicx]{revtex4-1}
\usepackage[utf8]{inputenc}
\usepackage{graphicx}% Include figure files
\usepackage{dcolumn}% Align table columns on decimal point
\usepackage{bm}% bold math
\usepackage{textcomp}  % for \textmu

\usepackage{wrapfig}

\begin{document}
%\preprint{}

\title{Single Molecule Nonlinearity in a Plasmonic Waveguide}

\author{Christian Sch\"orner}
\author{Markus Lippitz}
\email{markus.lippitz@uni-bayreuth.de}
\affiliation{Experimental Physics III, University of Bayreuth, Bayreuth, Germany}

%\date{\today}

\begin{abstract}
\setlength{\columnsep}{0pt}%
\begin{wrapfigure}{r}{0.5\textwidth}
  %\begin{center}
    \includegraphics[]{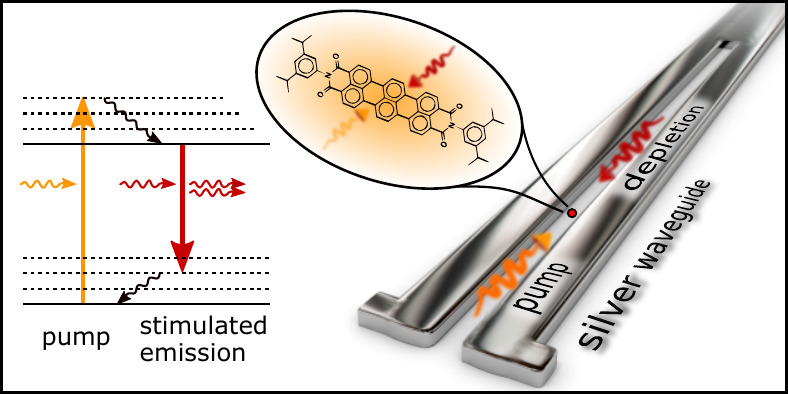}
  %\end{center}
\end{wrapfigure}
Plasmonic waveguides offer the unique possibility to confine light far below the diffraction limit. Past room temperature experiments of single quantum emitters coupled to such waveguides have mainly focused on efficient channelling of the spontaneous emission into waveguide modes after a linear optical excitation. However, only the simultaneous interaction of the emitter with multiple plasmonic fields would lead to functionality in a  plasmonic circuit. Here, we demonstrate the nonlinear optical interaction of a single molecule and propagating plasmons. An individual terrylene diimide (TDI) molecule is placed in the nanogap between two single-crystalline silver nanowires. A visible wavelength pump pulse and a red-shifted depletion pulse travel along the waveguide, leading to stimulated emission depletion (STED) in the observed fluorescence. The efficiency increases by up to a factor of 50 compared to far-field excitation. Our study thus demonstrates remote nonlinear four-wave mixing at a single molecule with propagating plasmons. It paves the way towards functional quantum plasmonic circuits and improved nonlinear single-molecule spectroscopy.
\end{abstract}

% insert suggested PACS numbers in braces on next line
\maketitle %\maketitle must follow title, authors, abstract and \pacs

\begin{figure}
\includegraphics{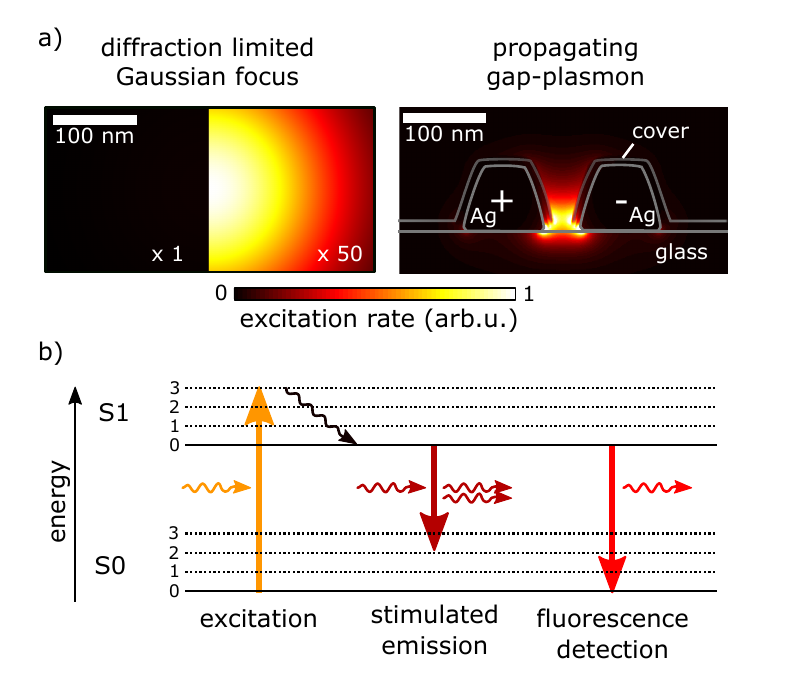}
\caption{Plasmonically enhanced nonlinearity of a single molecule. (a) A diffraction limited Gaussian focus (NA 1.4) has a much larger mode area than the gap-mode of a plasmonic waveguide. At the same transmitted power, the excitation rate of a molecule is thus up to 50 times larger. (b) The increased excitation probability enhances nonlinear optical effects such as stimulated emission depletion. After excitation and vibrational relaxation, a second laser pulse causes stimulated emission. The remaining population of the excited state is monitored by fluorescence.}
\label{fig:figure1}
\end{figure}

%introduction
Plasmonic waveguides confine light to deep subwavelength mode areas \cite{GramotnevNatPhot2010}. When a quantum emitter is placed in such a plasmonic mode, the emission rate is increased by the Purcell effect \cite{Tame2013,Huck2016}. Ideally, the emission into the waveguide mode overwhelms all other decay channels. Sources of single plasmons have been demonstrated in this way, both at room temperature and liquid helium temperature \cite{Akimov2007,Kolesov2010,Huck2011,Kumar2013,Li2015,Kress2015,Wu2017,Li2018,Kumar2018,Kumar2019,Grandi2019}, using various combinations of waveguides and emitters with suitable positioning procedures. While light absorption and fluorescence emission are linear optical properties of an emitter, all quantum emitters also show nonlinear optical effects, when interacting with more than one photon. Examples are two-photon absorption \cite{ZhangACSPhot2018}, stimulated emission \cite{Piatkowski2019}, and pump-probe spectroscopy in general. As these effects depend super-linearly on the total optical intensity, they should profit from plasmonic mode confinement and field enhancement (Figure \ref{fig:figure1}a). In this article, we present stimulated emission of a single plasmon by a single dye molecule coupled to a plasmonic waveguide.

Nonlinear light-matter interaction can be described by expanding the linear relation between incoming optical field $E$ and induced polarization $P$ to higher orders in the electric field \cite{Boyd2008}
\begin{equation}
    P = \epsilon_0 \left( \chi^{(1)} \, E + \chi^{(2)} \, E \, E + \chi^{(3)} \, E \, E \, E  + ... \right) \quad , \label{eq:chi1-3}
\end{equation}
where $ \chi^{(n)}$ is the susceptibility of order $n$. Stimulated emission and pump-probe spectroscopy in general are third-order nonlinear processes, governed by $\chi^{(3)}$. Three optical fields interact to produce a third-order nonlinear polarization $P^{(3)}$
\begin{equation}
    P^{(3)} = \epsilon_0  \chi^{(3)} \, E_{\text{pump}} \, E_{\text{pump}}^*  \, E_{\text{probe}}   = \epsilon_0  \chi^{(3)} \,| E_{\text{pump}} |^2  \, E_{\text{probe}} \label{eq:p3}
\end{equation}
This nonlinear polarization $P^{(3)}$ oscillates at the optical frequency of the probe beam. In a typical pump-probe experiment it is detected by interfering the radiation generated by the oscillating polarization with the probe beam. One thus detects a variation in the transmission of the probe beam, e.g.\ resulting from stimulated emission \cite{Min2009}. Alternatively, one can view the polarization as a quantum mechanical coherence between ground and excited state. Such a coherence can be transferred into a population by a fourth interaction with an optical field\cite{Tian:2003wxa}. The resulting population of the excited state is then detected by fluorescence emission. This scheme is used in  recent ultrafast nonlinear spectroscopy of a single molecule \cite{Liebel2018}, and in fluorescence-detected 2D-spectroscopy \cite{Mueller2018,Tiwari2018} in general.

To demonstrate a single-molecule nonlinearity in a plasmonic waveguide, we adopt this scheme. We use a pump pulse to excite the molecule from its ground state to a vibrational side band of the electronic excited state (Figure \ref{fig:figure1}b). On a (sub)picosecond time scale, the molecule relaxes to the vibrational ground state of the excited state. The second laser pulse is tuned such that it causes stimulated emission back into a high vibrational band of the ground state. The population after this interaction is monitored by fluorescence emission, which mainly occurs in the wavelength range just between the two laser pulses. The excitation scheme is identical to the pulse sequence used in stimulated emission depletion (STED) microscopy \cite{Kasper2010,Blom2017}. STED microscopy makes use of another nonlinearity, which goes beyond the $\chi^{(3)}$ description above (equations \ref{eq:chi1-3} and \ref{eq:p3}), namely that the stimulated emission saturates and the remaining fluorescence decreases exponentially with the depletion pulse energy \cite{Kastrup2004}.

\begin{figure}
\includegraphics{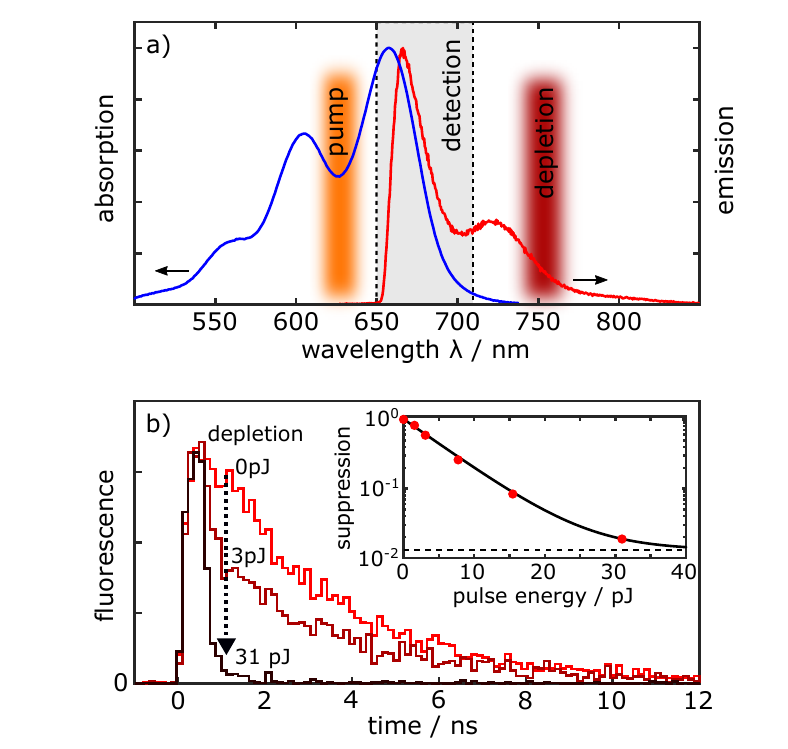}
\caption{Stimulated emission depletion of terrylene diimide (TDI) (a) Absorption and emission of a TDI ensemble in a poly-methyl-methacrylate (PMMA) polymer film. The chosen pump, depletion and detection wavelengths are indicated. (b) Fluorescence decay curves of a single TDI molecule as function of depletion pulse energy for a delay between pump and depletion pulse of 0.6\,ns. Inset: Fluorescence suppression factor $\eta$ (red dots) as a function of depletion pulse energy. An exponential fit with a background of about 1\,$\%$ is shown as black line.}
\label{fig:figure2}
\end{figure}
%

%Dye molecule as quantum emitter
In the experiment, we use terrylene diimide (TDI) molecules embedded in a thin poly-methyl-methacrylate (PMMA) polymer film as extremely stable quantum emitters \cite{Piwonski2015}. The spectral ranges of the two laser pulses and fluorescence detection are optimized for the TDI spectra (Figure \ref{fig:figure2}a). Both laser pulses are cut out of the same wide spectrum of a supercontinuum white light laser and have a temporal length of a few 10\,ps \cite{Silani2019}. We operate the laser at a repetition rate of 1.953~MHz and keep the sample under nitrogen atmosphere, both to reduce photobleaching of the dye. Additional to wide-field imaging, the setup allows accurate positioning of the excitation focus on the sample and scanning the relative position of depletion and detection focus independently of each other. More experimental details can be found in the supporting information (SI, Figures S1--S3).

\begin{figure*}
\includegraphics{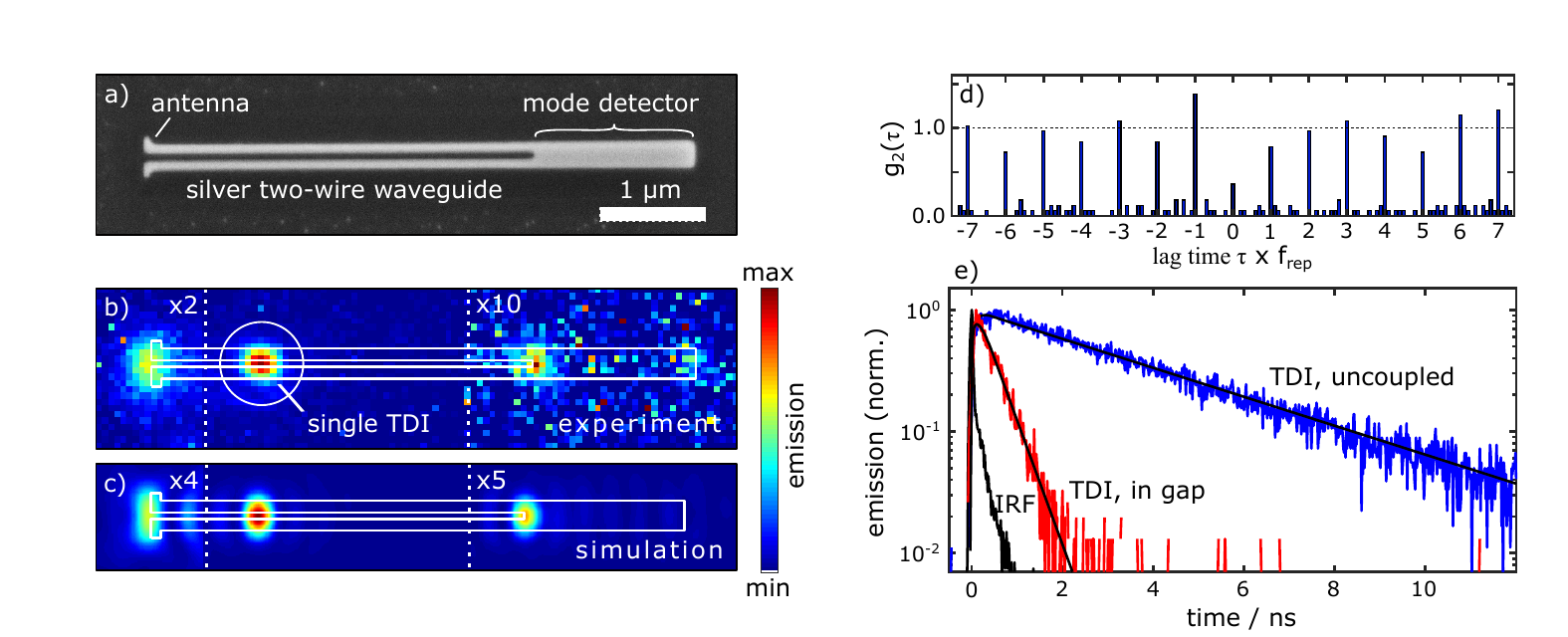}
\caption{(a) Scanning electron micrograph of the plasmonic two-wire waveguide with antenna and mode detector fabricated from single-crystalline silver. (b) Wide-field camera image of the fluorescence excited remotely via the antenna. The single emission spot along the gap is attributed to a single TDI molecule. (c) Simulated image of an electric dipole at the position of the TDI molecule. (d) Fluorescence auto-correlation of the TDI emission showing an antibunching dip at zero time delay ($\tau = 0$), proving single photon emission. (e) The fluorescence lifetime as measured by time-correlated single photon counting decreases from about 3.6\,ns in a thin PMMA film to values as low as 0.4\,ns when coupled to the waveguide. }
\label{fig:figure3}
\end{figure*}

Let us first discuss the reference case, stimulated emission depletion of single dye molecules in the diffraction-limited far-field focus of our microscope objective (NA 1.4). By time-correlated single photon counting we measure the arrival time of fluorescence photons after the pump pulse. At 0.6\,ns after the pump pulse, we shine in the depletion pulse of varying pulse energy (0--31~pJ). For low depletion pulse energy, we find the expected exponential decay with an excited state lifetime of about 3.6~ns on average (Figure \ref{fig:figure2}b). Increasing the depletion pulse energy, an abrupt drop appears in the fluorescence intensity, signaling the depopulation of the excited state by stimulated emission \cite{Rittweger2007,Bullock2010}. The inset of Figure \ref{fig:figure2}b shows the suppression factor $\eta$ by which the fluorescence decreases after the depletion pulse. It can be described by an exponential function of the depletion pulse energy\cite{Kastrup2004} with a characteristic pulse energy (6~pJ for the shown molecule). For a molecule well aligned with respect to a linearly polarized focus we estimate the saturating depletion pulse energy to be 1.2~pJ (Figure S4). Taking the spot size of the depletion focus (full width at half maximum: 270~nm, Figure S3) into account, this yields an absolute cross section for stimulated emission of about $1.8\times10^{-16}$cm$^2$, similar to values found for other dyes \cite{Kastrup2004,Rittweger2007}.

We now turn to coupling of single TDI molecules to a plasmonic two-wire transmission line (Figure \ref{fig:figure3}a), fabricated by focused ion-beam milling out of a 80\,nm high single crystalline silver flake\cite{Schoerner2019} (Figure S1). Single-crystalline silver, in general, represents the most preferred low-loss material at visible and near-IR frequencies for plasmonic applications\cite{ParkAM2012,WangNC2015,Zhang2019,Rodionov2019}. After milling, our waveguide structures are covered by a thin (5~nm) protection layer of $Al_2O_3$ to prevent chemical degradation\cite{Schoerner2019}. Both nanowires have a width of about 90\,nm, separated by a gap of about 60\,nm. A short dipole antenna at the left end of the waveguide allows efficient incoupling of photons from a Gaussian focus.
The two-wire transmission line supports two strongly confined plasmonic modes of different symmetry\cite{GeislerPRL2013, Schoerner2019}. 
The symmetric mode can be excited with light polarized parallel to the long axis of the waveguide and propagates until the far end of the waveguide with a low intensity in the gap. In contrast, the antisymmetric mode is excited by perpendicularly polarized light impinging on the antenna. This mode is characterized by opposing charges on the two wires and features the desired strong field confinement in the gap between the  nanowires, as already shown in Figure \ref{fig:figure1}a.  At the right end of the structure, the gap terminates earlier than the waveguide. This causes the antisymmetric mode to be reflected and scattered out to the far-field.
The right end of the transmission line acts thus as mode detector.

By spin coating a thin ($\sim$10\,nm) polymer film with a low concentration of TDI molecules on the waveguide sample, we find in some cases a single dye molecule in the gap between the two nanowires. Other dye molecules in close proximity to the structure are photobleached on purpose before further optical investigations. When exciting the antenna with a pump pulse (perpendicular polarization) the propagating antisymmetric mode in the gap excites the dye molecule. Imaging the fluorescence emission from the sample, we find one bright spot along the waveguide's gap, signaling the position of the dye molecule, and waveguide emission at the dipole antenna and the inner end of the mode detector, signaling emission from the antisymmetric mode (Figure \ref{fig:figure3}b). This emission behavior can be well reproduced by a dipole emitter in numerical simulations (Figure \ref{fig:figure3}c). For details see supplementary note 1.

The exact intensity ratio of the three emission spots depends on several factors. First, at the pump position on the antenna the experimental emission has an additional contribution of silver luminescence\cite{Peyser2001,KumarOptExpress2012}. Second, the waveguide geometry as well as the dipole position and orientation is idealized in the simulation, explaining a lower experimental signal at the mode detector. The emission spectra and the fluorescence images of the same structure before and after the photobleaching of the dye molecule, as well as fluorescence images with a different  position of the molecule along the gap are given in the SI (Figures S5 $\&$ S6).

The emission stems indeed from a single emitter, as the second-order correlation function shows clear antibunching at a lag $\tau=0$, with a value $g^{(2)}(\tau=0) = 0.36$ well below 0.5 (Figure \ref{fig:figure3}d). The  non-zero value is fully explained by coincidences involving the dark count rate of the detectors and weak luminescence from the silver structure which has been quantified after the single-step photobleaching of the dye (Figure S5).

The fluorescence lifetime reduces from about 3.6~ns in the uncoupled case to values as low as 0.4\,ns when placed in the gap of the waveguide (Figure \ref{fig:figure3}e, for statistics see Figure S6). This lifetime reduction of about one order of magnitude is also found for $WSe_2$ quantum emitters near a gold two-wire waveguide \cite{BlauthNanoLett2018} and by numerical simulations \cite{Chen2010}. For an electric dipole at the position of the molecule along the gap (Figure \ref{fig:figure3}c) the total decay rate is found to be 11.7 times higher than in vacuum or 9.6 times higher compared to the substrate only case, further supporting our experimental findings. For different dipole positions along the gap, decay rate enhancements above 20 can be found in simulations due to a position-dependent Purcell factor.

\begin{figure}
\includegraphics{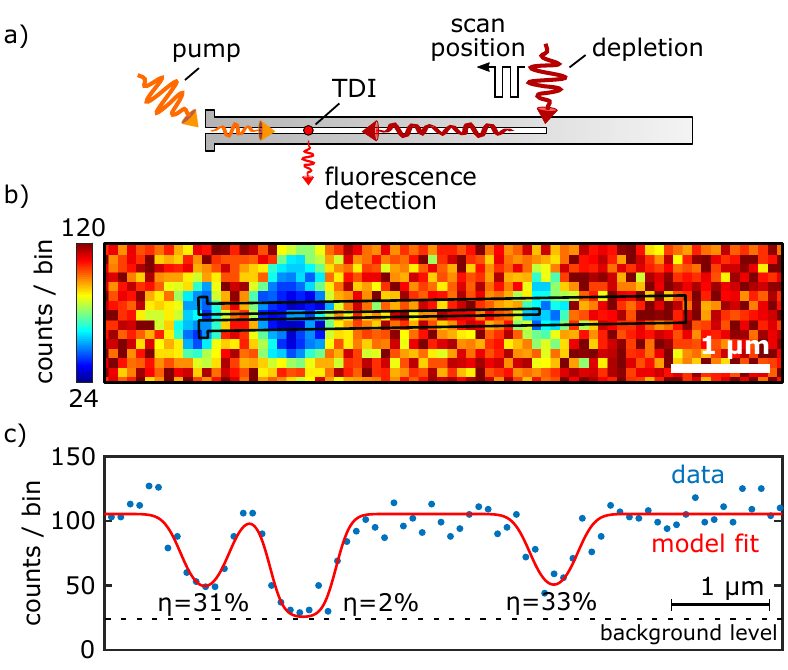}
\caption{Nonlinear interaction of a single dye molecule with propagating gap-plasmons. (a) The molecule is excited via a pump pulse focused on the antenna. The depletion pulse focus is raster-scanned over the sample while the fluorescence is detected. (b) Resulting fluorescence intensity map. (c) Profile of the data in (b) close to center of the waveguide and fit of the model to the data.}
\label{fig:figure4}
\end{figure}

To force the same molecule in the nanogap to undergo stimulated emission we now switch on the depletion pulses. We scan the position of the depletion focus (perpendicular polarization) over the structure, while keeping the pump focused on the antenna and monitoring fluorescence emission into free space at the position of the molecule (Figure \ref{fig:figure4}a). The delay between the optical pulses is reduced to 50\,ps, well below the lifetime of the molecule, to optimize the population available for  stimulated emission. The resulting fluorescence intensity map (Figure \ref{fig:figure4}b) is maximal whenever the depletion pulse does not reach the molecule and the fluorescence emission is not disturbed. When the depletion pulse is focused directly on the molecule, the nonlinear interaction results in stimulated emission and the fluorescence emission is suppressed down to the background level of about 24 counts per bin. But also when the depletion pulse is focused on the antenna or the inner end of the mode detector, we find fluorescence suppression, down to about 30\,\% of the maximum value. At those latter positions the depletion pulse can launch the strongly confined antisymmetric waveguide mode (Figure S7) which propagates toward the molecule and induces stimulated emission. We thus have detected the stimulated emission of single plasmons in the waveguide by a single terrylene diimide molecule. The strong fluorescence suppression demonstrates that we already enter the saturation regime of the stimulated emission. In contrast, for a parallel polarization of the depletion pulse launching the symmetric mode with a weak intensity in the gap \cite{Schoerner2019}, no stimulated emission occurs.

Despite the shorter distance of the molecule to the antenna, the fluorescence suppression after a depletion pulse via the antenna and via the mode detector is nearly equal (Figure \ref{fig:figure4}b). In numerical simulations we find, besides similar incoupling efficiencies at both positions, a small contribution of a short-ranged leaky higher order mode\cite{Schoerner2019} close to the antenna. Its superposition with the antisymmetric mode slightly reduces the depletion pulse intensity at the position of the molecule. 

A horizontal cross section through the center of the map in Figure \ref{fig:figure4}b can be well described by a model of Gaussian spots at the positions of antenna, molecule, and mode detector, taken as input to the exponential suppression (Figure \ref{fig:figure4}c, supplementary note 2).

The pulse energy of the linearly polarized depletion pulse focused to the inner end of the mode detector was 6\,pJ. Taking the experimental full transmission of 0.12\,\% into account (Figure S7) and correcting for outcoupling/collection efficiency by the antenna (simulated to be about 28\,\%) and the propagation loss to the molecule (about 40\,\%, Figure S8), we arrive at about 42\,fJ depletion pulse energy that pass the molecule in the antisymmetric plasmonic waveguide mode. As this leads to a fluorescence reduction down to about 33\,\%, the saturating depletion energy of the antisymmetric mode can be estimated to be about 38\,fJ, a factor of about 30 lower than the optimal 1.2\,pJ in the free-space reference experiment above. Numerical simulations show that the experimental value of about 30 should reach about 50 for a perfectly positioned and orientated molecule in the gap (Figure \ref{fig:figure1}a). 

To summarize, we demonstrated coupling of a single dye molecule to a plasmonic waveguide operating in the visible spectral range. A clear antibunching dip signals the presence of only one emitter, with a fluorescence lifetime reduced by about one order of magnitude compared to the uncoupled case. Stimulated emission of a single plasmon via a plasmonic mode is detected by its influence on the free-space fluorescence emission. The required pulse energy is reduced by about a factor of 30 compared to a free space Gaussian focus. This nonlinear plasmon-plasmon interaction at a single molecule opens up new opportunities for nanooptics and plasmonics. The plasmonic waveguide could be used to simplify ultrafast nonlinear spectroscopy of single molecules \cite{Liebel2018}, as the reduced excited state lifetime reduces photobleaching\cite{Cang2013}. For this purpose, the laser pulses need to be shortened from at the moment several 10\,ps to a few 10\,fs. The linear chirp imposed on ultrafast  pulses after propagating 1\,$\mu$m of the current waveguide equals approximately that of 100\,$\mu$m glass, which is routinely pre-compensated. Future engineering of mode area and transmission of the waveguide would also give access to the coherent nonlinearity of single emitters \cite{Faez2014,Turschmann2017,Turschmann2019,Javadi2015}, even at room temperature. In such experiments, not the fluorescence but the coherent transient transmission would be detected but requires considering the waveguides' own nonlinearity\cite{Li2019}. Furthermore, the flexible fabrication technique by focused ion beam milling from single-crystalline silver easily allows to add further plasmonic circuit elements like beam splitters, routers, and interferometers. Together with deterministic positioning methods of molecules \cite{Hail2019,Kewes:2016jn} this will open the field of integrated quantum plasmonic circuitry.

\begin{acknowledgments}
The authors thank Patrick Kn\"odler for operating our FIB-milling patterns at BGI Bayreuth and SEM imaging at BPI Bayreuth. We thank Philipp Ramming for help with the ALD machine. C.S. gratefully acknowledges financial support from the Deutsche Forschungsgemeinschaft (GRK1640).
\end{acknowledgments}

% Create the reference section using BibTeX:
%merlin.mbs aipnum4-1.bst 2010-07-25 4.21a (PWD, AO, DPC) hacked
%Control: key (0)
%Control: author (8) initials jnrlst
%Control: editor formatted (1) identically to author
%Control: production of article title (0) allowed
%Control: page (1) range
%Control: year (1) truncated
%Control: production of eprint (0) enabled

%

\end{document}

\ohead[]{}
\frenchspacing
\raggedbottom
\selectlanguage{american} % american ngerman
%\renewcommand*{\bibname}{new name}
%\setbibpreamble{}
\pagenumbering{roman}
\pagestyle{plain}
%********************************************************************
% Frontmatter
%*******************************************************

\ofoot[]{}% keine Seitenzahl mehr außen (o = near outer margin)
\cfoot[\pagemark]{\pagemark}% Seitenzahl (c = centered)

\renewcommand{\thepage}{S\arabic{page}} 
\begin{center}

        \hfill

		{\normalsize{\spacedallcaps{Supporting Information}}}\\

     \vspace{2cm}

		{\LARGE Singe Molecule Nonlinearity in a Plasmonic Waveguide}

		\vspace{2cm}
		Christian Sch\"orner$^\dagger$ and Markus Lippitz$^{\dagger,*}$\\
		\vspace{0.5cm}
		$\dagger$ \, Experimental Physics III, University of Bayreuth, Universit\"atsstr.\ 30, 95440 Bayreuth, Germany\\
		\end{center} 
		
		\vspace{2cm}
		Corresponding Author\\
		\ * Email: markus.lippitz@uni-bayreuth.de
		
		\vspace{1cm}
		\textbf{table of content}\\
		
		Figure S1: Sample fabrication\\
		
		Figure S2: Optical setup\\
		
		Figure S3: Pump and depletion spot size\\
		
		Figure S4: Depletion statistics of TDI\\
		
		Figure S5: Lifetime histogram of TDI\\
		
		Figure S6: Quantifying the background emission\\
		
		Figure S7: Propagation of the depletion pulses\\
		
		Figure S8: Propagation loss with PMMA cover\\
		
		Supplementary note 1: numerical simulations\\
		
		Supplementary note 2: model for fluorescence suppression

\pagestyle{scrheadings}

%*******************************************************
\let\cleardoublepage\clearpage
\cleardoublepage\pagenumbering{arabic}

\cleardoublepage

	\renewcommand{\thepage}{S\arabic{page}} 
	\renewcommand{\thefigure}{S\arabic{figure}}
	\setcounter{page}{2}
	
%************************************************
\chapter*{Sample Fabrication}
\ofoot[]{}% keine Seitenzahl mehr außen (o = near outer margin)
\cfoot[\pagemark]{\pagemark}% Seitenzahl (c = centered)
%************************************************
\vspace*{-1cm}
\begin{figure}[ht!]
				\centerline{\includegraphics[]{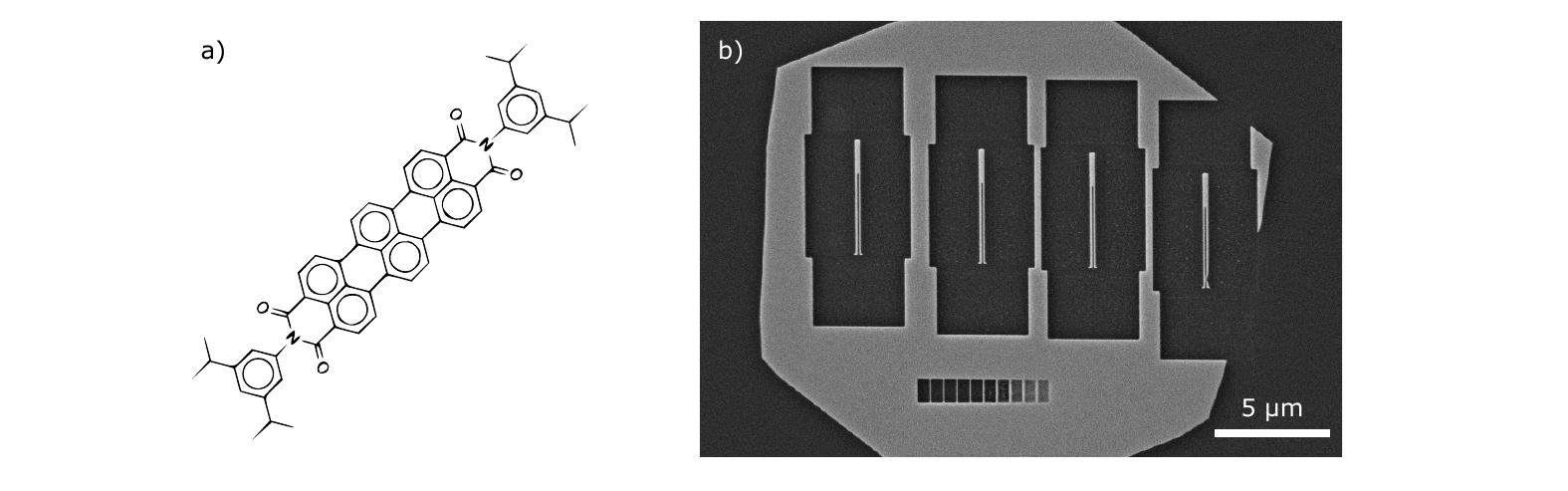}}
				\caption{\label{fig:S1} (a) Terrylene diimide was bought from KU dyes ApS, Copenhagen. As matrix material we use a PMMA resist (ARP671.015, Allresist) that was further diluted to 0.5\,\% solid content. The solution was doped with TDI at a very low $<10^{-9}\,$mol/L concentration and spin coated on the waveguide sample at 1000 rotations per minute resulting in a film thickness of only about 10\,nm. For ensemble measurements, a higher concentration of TDI was used. (b) The waveguide fabrication is performed by focused ion beam milling (Scios, FEI company) using Ga ions at 30\,kV acceleration voltage and 49\,pA ion current. The waveguide circuits are milled in a thin silver flake (height 80\,nm) on glass covered with a thin conductive layer (Electra 92, Allresist). The conductive layer is easily removed after milling via DI water. As a result, silver waveguides are remaining, well separated ($>$2\,\textmu m) from the silver flake for optical measurements. Furthermore, the silver nanostructures are encapsulated with a thin layer of $Al_2O_3$ deposited by atomic layer deposition (Savannah, Ultratech). We use 60\,cycles of water and TMA precursor-pulses at 100$^\circ C$ deposition temperature, yielding a layer of about 5\,nm in thickness. For the optical measurements in this work the three nanostructures on the left side (same fabrication parameters) have been used. The structure at the right side was excluded due to a fabrication imperfection.}
\end{figure}

%************************************************
\chapter*{Optical setup}
\ofoot[]{}% keine Seitenzahl mehr außen (o = near outer margin)
\cfoot[\pagemark]{\pagemark}% Seitenzahl (c = centered)
%************************************************
\vspace*{-1cm}
\begin{figure}[ht!]
				\centerline{\includegraphics[]{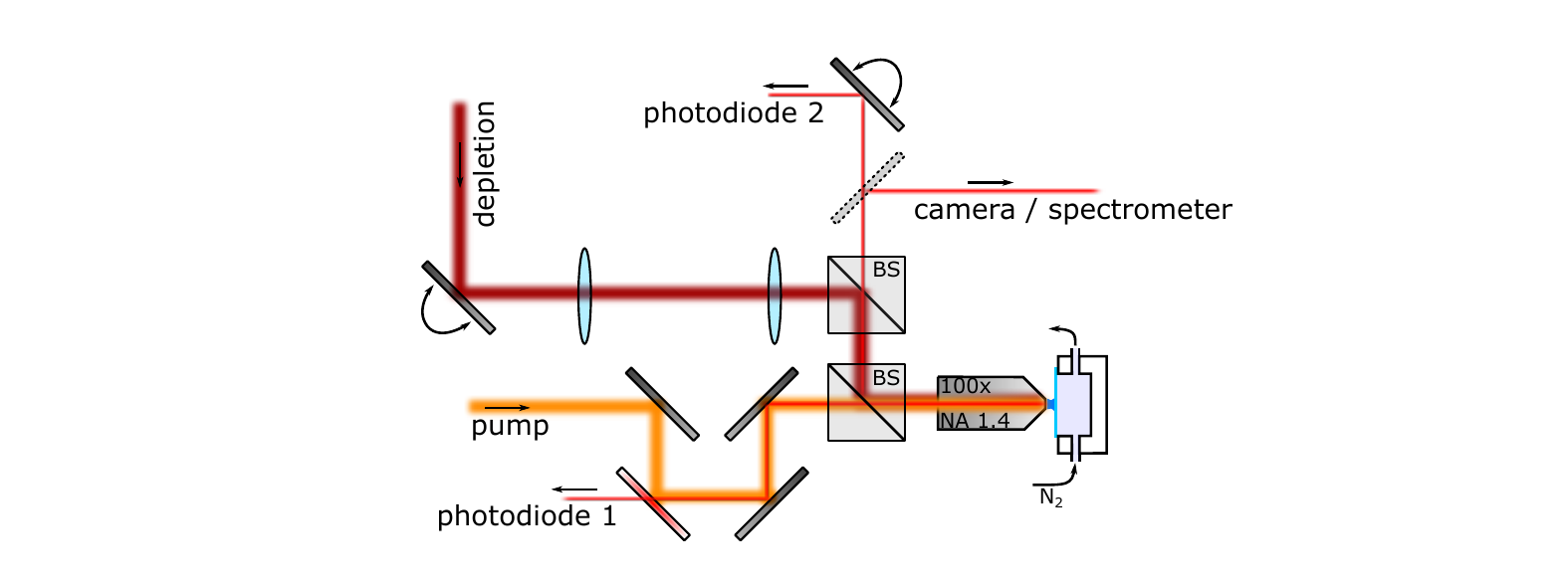}}
				\caption{\label{fig:S2} Simplified sketch of the setup. As light source we use a supercontinuum white light laser (SuperK EXTREME EXR-15, NKT Photonics) equipped with a tunable single line filter (SuperK Varia, NKT Photonics). The output spectrum (610-770\,nm) is split into a pump and depletion path via a dichroic mirror (zt 635RDC, AHF Analysentechnik). Bandpass filters define the final spectral ranges of the pump (BP 625/30, AHF) and depletion path (2x BP 769/41, AHF). Polarizers (LPVIS100-MP and WP25M UB, Thorlabs) can be inserted to define a linear polarization. The two light paths are superimposed in front of the objective by two 50:50 beamsplitters (BS016 and BS013, Thorlabs). The focus position of the depletion pulses on the sample can be scanned via a piezo mirror platform (S-334.2SL, Physik Instrumente) equipped with a piezo controller (E-501.00, Physik Instrumente), followed by a 4f-system consisting of two lenses (250\,mm and 300\,mm focal length). The setup offers two detection paths. Light in path 1 passes the 50:50 beamsplitter, a dichroic mirror (HC BS650, AHF) and dielectric longpass (LP635, AHF) and shortpass (SP710, SP730) filters and is finally focused to avalanche photodiode 1 (MPD PDM Series, Picoquant) via a 150\,mm tube lens. The light in path 2 is reflected by the first beamsplitter, transmitted by the second beamsplitter and then several detectors can be selected by flip mirrors. First, the light can be guided to an avalanche photodiode 2 (MPD PDM Series, Picoquant) via a scanning mirror (PSH 25, Piezosystem Jena) equipped with a piezo controller (d-Drive, Piezosystem Jena) and a tube lens (150\,mm focal length). Second, the light in path 2 can be guided toward the CMOS camera (Zyla 4.2, Andor) with a tube lens of 200\,mm. Third, the light in path 2 can be guided to a spectrometer (Isoplane 160 with ProEM EMCCD camera, Princeton Instruments). For fluorescence measurements via path 2 a longpass filter (LP647, AHF) is flipped in the detection path. The fluorescence in case of the two-pulse experiment is detected via the photodiode in path 1. Both photodiodes have a detection spot size with about 600\,nm diameter in the sample plane. The pulse energies of the pump pulses have been about 1\,pJ. All pulse energies are given in the sample plane after passing the objective. Typically, the integration time was 2\,s for the imaging camera and 10\,s for the spectrometer camera. The time bin of figure 4b in the main manuscript was 80\,ms. }
\end{figure}

%************************************************
\chapter*{Pump and depletion spot size}
\ofoot[]{}% keine Seitenzahl mehr außen (o = near outer margin)
\cfoot[\pagemark]{\pagemark}% Seitenzahl (c = centered)
%************************************************
\vspace*{-1cm}
\begin{figure}[ht!]
				\centerline{\includegraphics[]{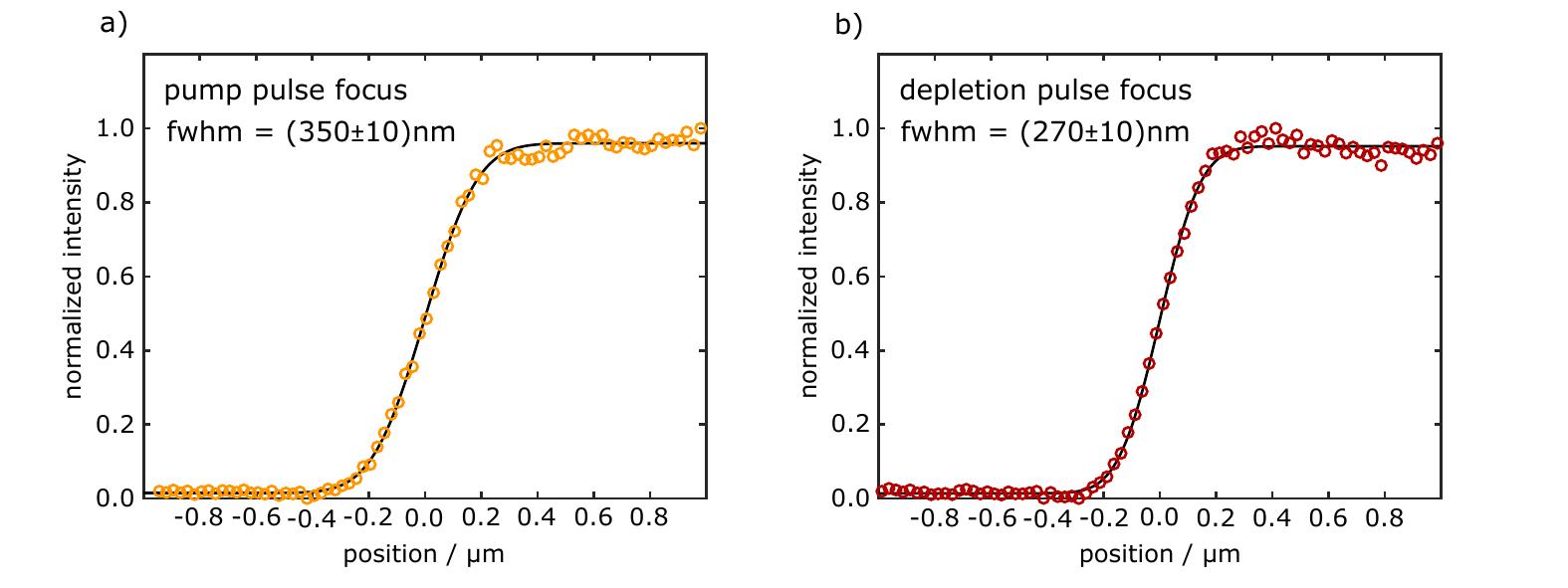}}
				\caption{\label{fig:S3} Laser spot sizes measured by scanning the sharp edge of a silver flake across each laser focus and recording the reflected intensity. The experimental data (averaged over 5 consecutive scans) is shown as colored circles while the fit of an error-function is shown as black solid line. The polarization is set parallel to the edge to avoid the excitation of surface plasmons. The full width at half maximum for the pump-pulse (a) and depletion-pulse (b) focus is calculated from the width of the error-function.}
\end{figure}

%************************************************
\chapter*{Depletion statistics of TDI}
\ofoot[]{}% keine Seitenzahl mehr außen (o = near outer margin)
\cfoot[\pagemark]{\pagemark}% Seitenzahl (c = centered)
%************************************************
\vspace*{-1cm}
\begin{figure}[ht!]
				\centerline{\includegraphics[]{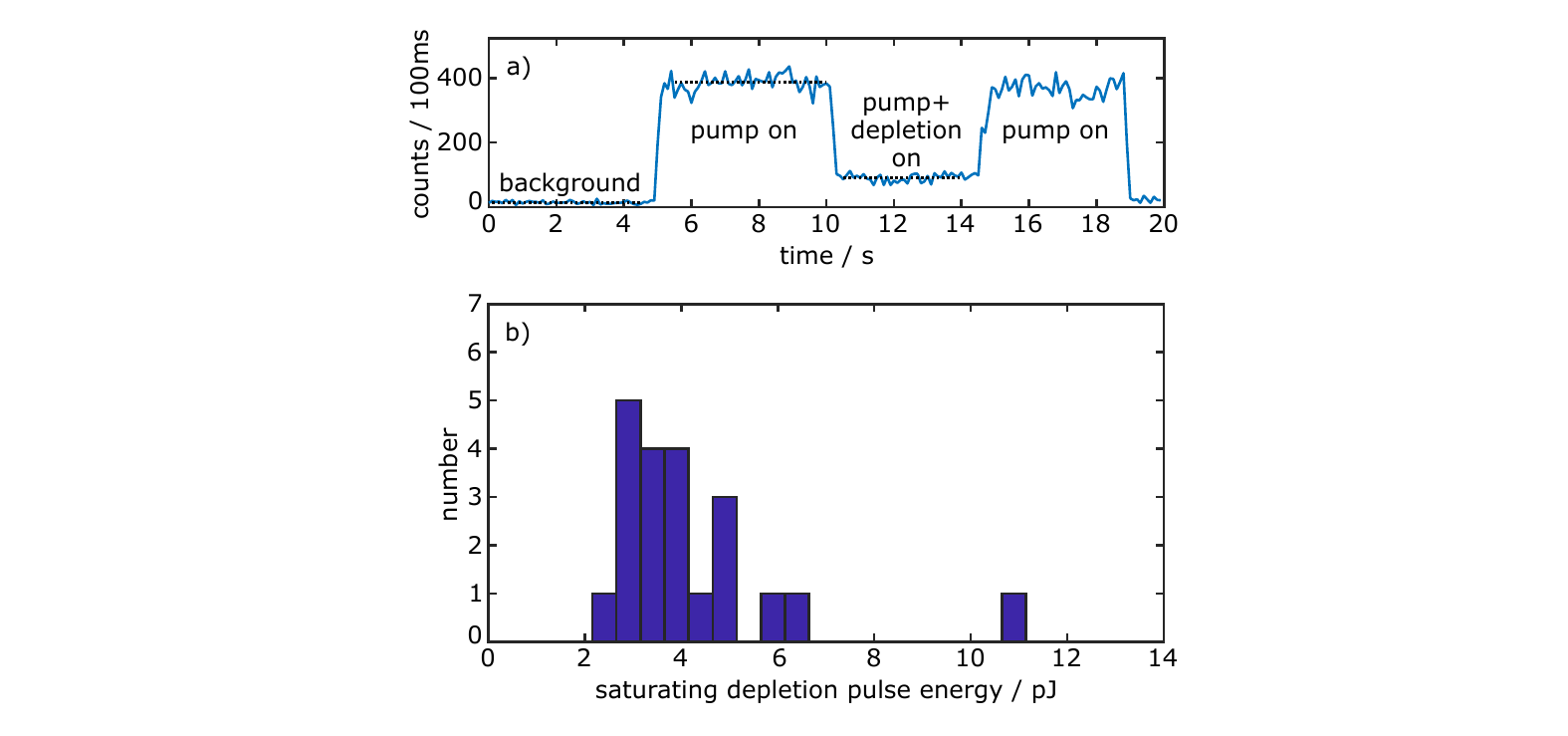}}
				\caption{\label{fig:S4} (a) Time trace of the emission of a single TDI molecule in PMMA. The pump is switched on at 5\,s and the depletion pulses ($E_{depletion} = 3.9$\,pJ) are added during about 10--15\,s. At 19\,s the pump is turned off. The fluorescence suppression $\eta$ is calculated using the three relevant intensity levels (black dashed). Subsequently, the saturating depletion pulse energy is calculated according to $E_{sat} = -E_{depletion} / ln(\eta)$. (b) Histogram of the saturation depletion pulse energy for different single TDI molecules. We attribute the variation of the saturation pulse energy from molecule to molecule mainly to the orientation of the TDI in the PMMA film. Pump and depletion pulses are unpolarized for these measurements. In case of a linear depletion pulse polarization well aligned with the molecular transition dipole moment a saturation depletion energy of 1.2\,pJ is estimated.}
\end{figure}

%************************************************
\chapter*{Quantifying the background emission}
\ofoot[]{}% keine Seitenzahl mehr außen (o = near outer margin)
\cfoot[\pagemark]{\pagemark}% Seitenzahl (c = centered)
%************************************************
\vspace*{-1cm}
\begin{figure}[ht!]
				\centerline{\includegraphics[]{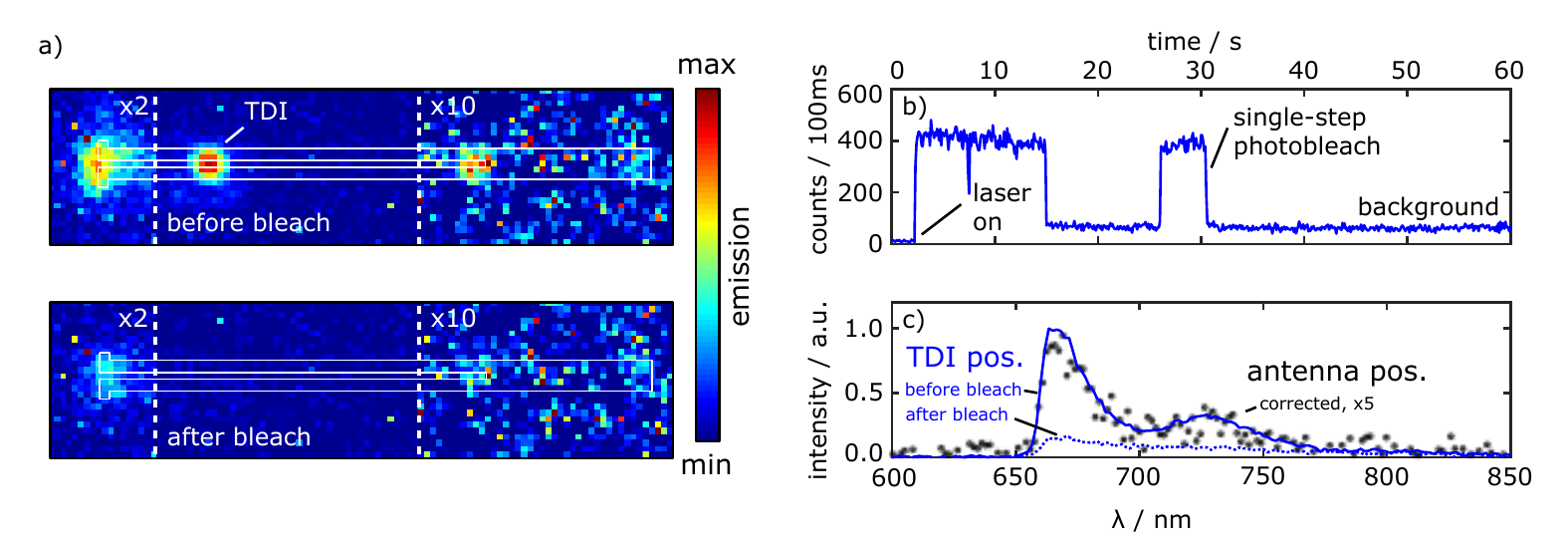}}
				\caption{\label{fig:S5} (a) Fluorescence image of the waveguide structure before (top) and after (bottom) photobleaching of the TDI. The pump-pulses are focused on the antenna with perpendicular polarization. After photobleaching, the emission spot along the gap, as well as most emission at the inner end of the mode detector, vanishes. Furthermore, the emission at the excitation position (antenna) is strongly reduced. (b) Time trace of the emission at the TDI position demonstrating a single-step photobleaching of the coupled TDI. (c) Spectrum recorded at the TDI position (blue, solid) before the photobleaching event resembling the emission spectrum of TDI molecules with a vibronic progression. After the photobleaching a weak and unstructured broad spectrum (blue, dotted) is detected at that position. The spectrum at the antenna - corrected by the background emission collected after the photobleaching - is shown as black dots (multiplied by a factor of 5). We attribute the weak background emission, which is present at the excitation position even after photobleaching of the TDI, to luminescence of silver.}
\end{figure}

%************************************************
\chapter*{Lifetime histogram of TDI}
\ofoot[]{}% keine Seitenzahl mehr außen (o = near outer margin)
\cfoot[\pagemark]{\pagemark}% Seitenzahl (c = centered)
%************************************************
\vspace*{-1cm}
\begin{figure}[ht!]
				\centerline{\includegraphics[]{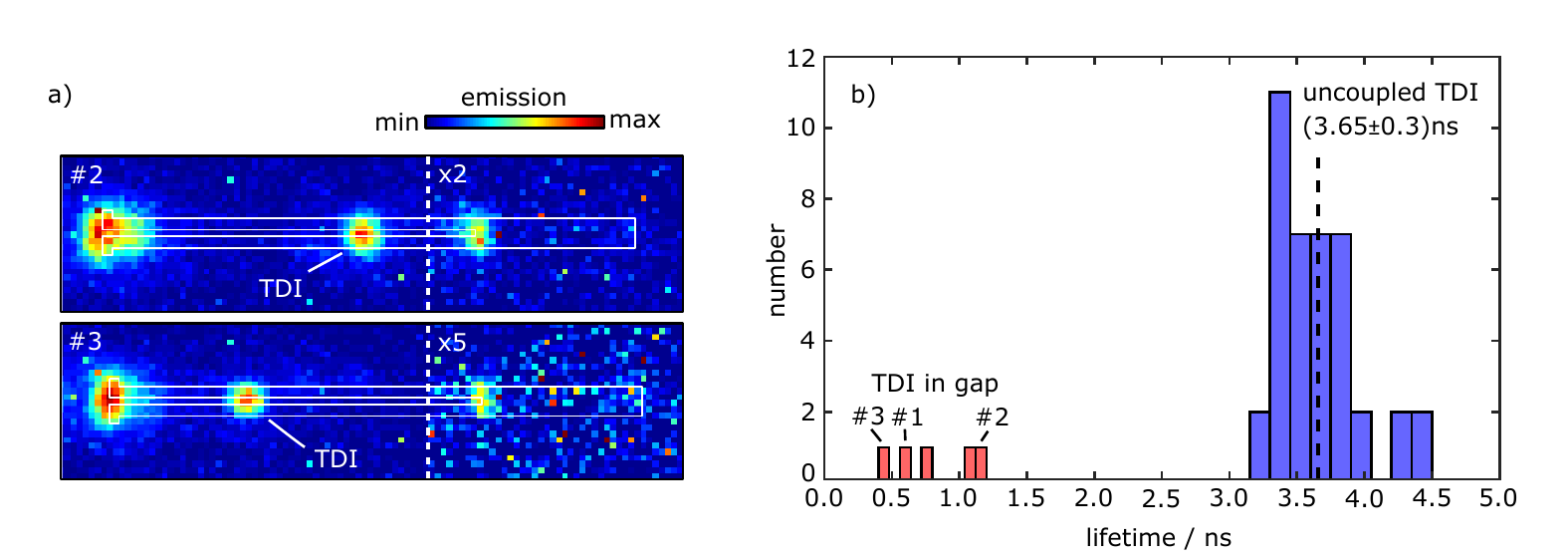}}
				\caption{\label{fig:S6} (a) Fluorescence images of additional exemplary datasets $\#2$ and $\#$3 of a single TDI molecule in the gap of the waveguide structure. The fluorescence image of dataset $\#$1 is shown in the main manuscript and Figure S5. The pump pulse focus is positioned on the antenna with perpendicular polarization in all cases, which results in a contribution of silver luminescence at this position (c.f.\ Figure S5). (b) Lifetime histogram of TDI in the uncoupled case (blue) and in the gap of the waveguide structure (red) detected directly at the TDI position with a detection spot size of 600\,nm in diameter.}
\end{figure}

%************************************************
\chapter*{Propagation of the depletion pulses}
\ofoot[]{}% keine Seitenzahl mehr außen (o = near outer margin)
\cfoot[\pagemark]{\pagemark}% Seitenzahl (c = centered)
%************************************************
\vspace*{-1cm}
\begin{figure}[ht!]
				\centerline{\includegraphics[]{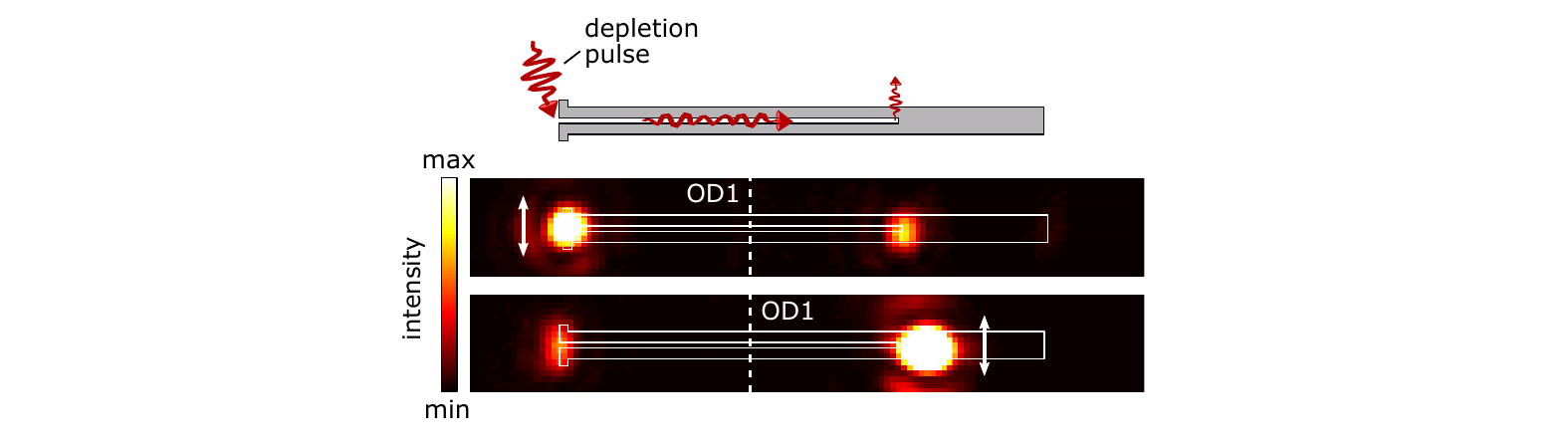}}
				\caption{\label{fig:S7} Far-field imaging of the propagation of the depletion pulses via gap-plasmons (antisymmetric mode) upon focusing on the antenna (top) and near end of the mode detector (bottom). Without PMMA cover the transmission is 0.3\%, whereas with PMMA cover the transmission reaches 0.12\%, likely reduced due to the increased refractive index in the gap. We define the transmission as the power scattered from the waveguide end in the collection angle of the objective relative to the applied power at the excitation spot. The latter we determine from its reflection at a cleaned substrate surface, corrected by the surface reflectivity. In total, this transmission includes the incoupling, propagation, and out-coupling/collection efficiency.}
\end{figure}

%************************************************
\chapter*{Propagation loss with PMMA cover}
\ofoot[]{}% keine Seitenzahl mehr außen (o = near outer margin)
\cfoot[\pagemark]{\pagemark}% Seitenzahl (c = centered)
%************************************************
\vspace*{-1cm}
\begin{figure}[ht!]
				\centerline{\includegraphics[]{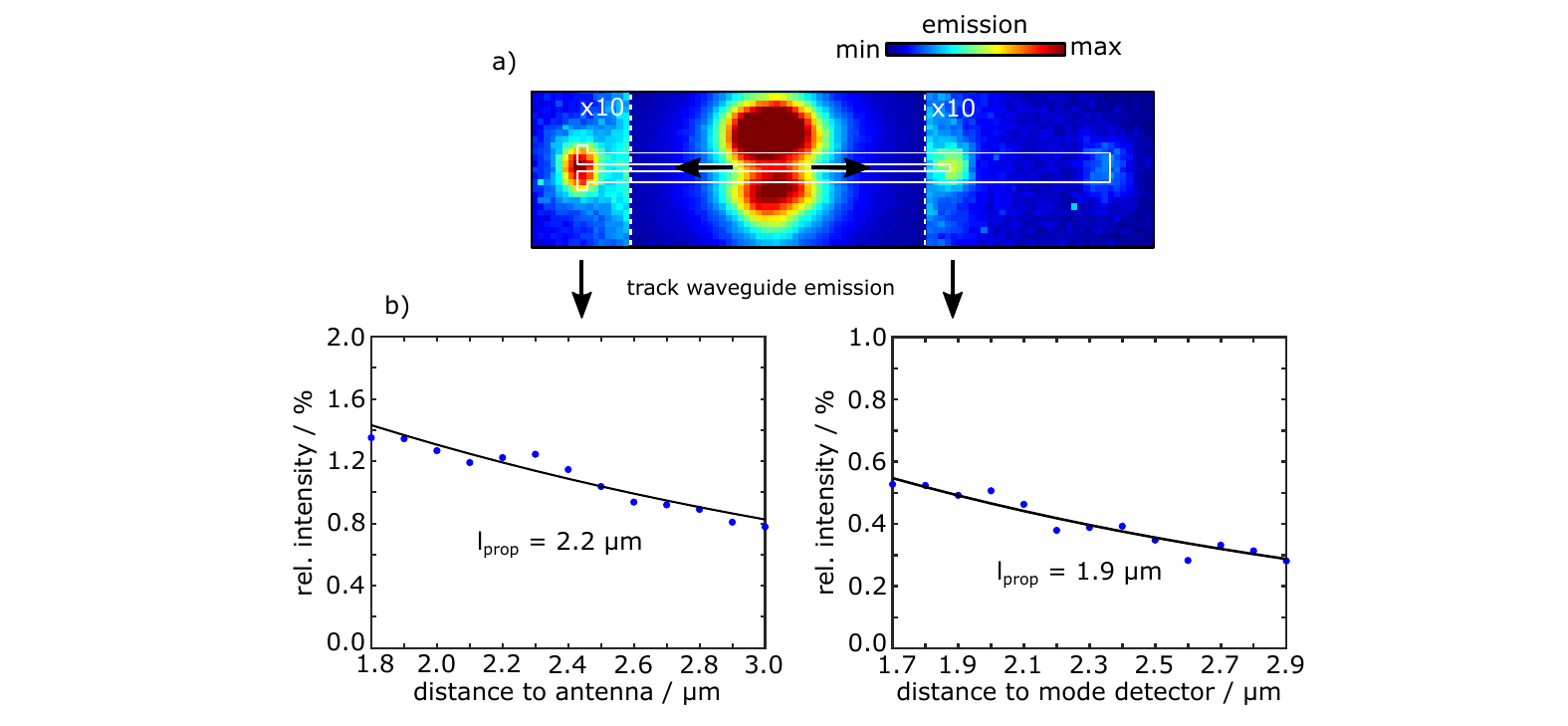}}
				\caption{\label{fig:S8} (a) Fluorescence image of an ensemble of TDI molecules in a PMMA layer on the waveguide structure with a pump focus at the center and polarized perpendicular to the long nanowires' axis. (b) The waveguide emission at the antenna (left) and inner end of the mode detector (right) normalized to the emission at the excitation position. The propagation loss is determined by a single exponential fit, yielding a value of about 2\,$\mu$m with the PMMA cover. Without the PMMA cover the propagation length is expected to be higher as the transmission of the structure is increased by a factor of 2.5 (see Figure S7).}
\end{figure}

%************************************************
\chapter*{Supplementary note 1: numerical simulations}
\ofoot[]{}% keine Seitenzahl mehr außen (o = near outer margin)
\cfoot[\pagemark]{\pagemark}% Seitenzahl (c = centered)
%************************************************

Electromagnetic near-fields are calculated using the commercial software package Comsol Multiphysics. The refractive index for silver is taken from Johnson and Christy [1]. The air and substrate are modeled by the refractive index 1.0 and 1.5, respectively. The near-to-far-field transformation was computed using a method based on reciprocity arguments using a freely available software package [2]. For imaging, the far-field is first refracted at the objective and tube lens taking the finite numerical aperture of the objective into account. The propagation to the image plane is subsequently performed by a Fourier transform [3]. The far-field imaging simulation is performed with an in-plane dipole ($\lambda =680$\,nm) oriented across the gap in 20\,nm height above the substrate. The thin cover (ALD and PMMA) layer is neglected for numerical simplicity of the full 3-dimensional model.

%************************************************
\chapter*{Supplementary note 2: model for fluorescence suppression}
\ofoot[]{}% keine Seitenzahl mehr außen (o = near outer margin)
\cfoot[\pagemark]{\pagemark}% Seitenzahl (c = centered)
%************************************************

The model of the intensity cross-section $I$ (units: counts per 80\,ms time bin) in figure 4c of the main manuscript is given by:
\begin{equation}
I = I_{bg} + I_{TDI} \sum_n \exp \left( - a_n G(x - x_{n}) \right) \quad \text{.}
\end{equation}
The background level is fixed to $I_{bg} = 24$ counts, determined by the dark counts of the detector and weak luminescence of the silver. $I_{TDI}$ represents the average undisturbed TDI fluorescence counts. Normalized Gaussian spots $G(x)$ at the positions $x_n$ of antenna, molecule, and mode detector, are taken as input to the exponential suppression. $a_n$ describes the saturation factor by which the depletion pulse fluence applied to the molecule exceeds the saturating pulse fluence.\\
An optimization of all non-fixed variables is performed, resulting in $I_{TDI} = 81$ and the suppression factors $\eta_n = exp(-a_n)$ given in figure 4c of the main manuscript. The intrinsic width of the Gaussian is about 450\,nm, probably broadened compared to the focus size (Figure S3) due to the finite size of the silver nanostructure and finite pixel size. The exponential function further broadens the resulting width of the fluorescence suppression dip. This effect is most pronounced in the saturation regime, c.f.\ the position of the molecule in figure 4c of the main manuscript.

\manualmark
\markboth{\spacedlowsmallcaps{\bibname}}{\spacedlowsmallcaps{\bibname}} % work-around to have small caps also
%\phantomsection 
\refstepcounter{dummy}
\addtocontents{toc}{\protect\vspace{\beforebibskip}} % to have the bib a bit from the rest in the toc
\addcontentsline{toc}{chapter}{\tocEntry{\bibname}}
\label{app:bibliography}

%\printbibliography

%merlin.mbs aipnum4-1.bst 2010-07-25 4.21a (PWD, AO, DPC) hacked
%Control: key (0)
%Control: author (8) initials jnrlst
%Control: editor formatted (1) identically to author
%Control: production of article title (0) allowed
%Control: page (1) range
%Control: year (1) truncated
%Control: production of eprint (0) enabled
%